\documentclass[aps,prb,twocolumn]{revtex4}
\usepackage{graphicx}

\begin{document}

\title{Plasmonic enhancement of fluorescence and Raman scattering by metal nanotips}

\author{N.\ I.\ Cade}
\email{nicholas.cade@kcl.ac.uk}
\author{F.\ Culfaz}
\author{L.\ Eligal}
\author{T.\ Ritman-Meer}
\author{F-M.\ Huang}
\author{F.\ Festy}
\author{D.\ Richards}
\affiliation{Department of Physics, King's College London, Strand, London WC2R 2LS, UK}


      \begin{abstract}
We report modifications to the optical properties of fluorophores in the vicinity of noble metal nanotips. The fluorescence from small clusters of quantum dots has been imaged using an apertureless scanning near-field optical microscope. When a sharp gold tip is brought close to the sample surface, a strong distance-dependent enhancement of the quantum dot fluorescence is observed, leading to a simultaneous increase in optical resolution.  These results are consistent with simulations of the electric field and fluorescence enhancement near plasmonic nanostructures.
Highly ordered periodic arrays of silver nanotips have been fabricated by nanosphere lithography. Using fluorescence lifetime imaging microscopy, we have created high resolution spatial maps of the lifetime components of vicinal fluorophores; these show an order of magnitude increase in decay rate from a localized volume around the nanotips, resulting in a commensurate enhancement in the fluorescence emission intensity. Spatial maps of the Raman scattering signal from molecules on the nanotips shows an enhancement of more than 5 orders of magnitude.
\end{abstract}

\keywords{Scanning near-field optical microscopy, ASNOM, Fluorescence lifetime imaging, FLIM, plasmon, nanoparticle, Raman, SERS}

\maketitle

\section{Introduction}
\label{sec:intro}

Fluorescence spectroscopy is a powerful tool in the study of biological samples; however, conventional imaging techniques suffer from limited sensitivity at the low fluorophore concentrations typically employed, and have a spatial resolution generally much coarser than the size of the structures of interest. Optical imaging with a resolution below the diffraction limit has been made possible with the development of scanning near-field optical microscopy (SNOM).\cite{hecht00} This has enabled measurements of fluorescence from single molecules\cite{frey04} and nanocrystals\cite{protasenko04,gerton04} with resolutions down to about 10 nm. One implementation of this is apertureless-SNOM (ASNOM), in which a sharp tip is used to modify the local electric field illuminating a sample surface.\cite{patane04}

The presence of a metallic structure can dramatically alter the emission properties of a locally situated fluorophore:\cite{barnes98} The optical field in the vicinity of a metal tip is strongly enhanced due to both the resonant excitation of localized surface plasmons and the lightning rod effect of highly curved metal surfaces. Furthermore, metallic structures can also  modify  both the radiative and the nonradiative decay rates for a molecule, resulting in enhanced fluorescence emission and greater photostability by reducing the excited state lifetime.\cite{lakowicz02,dulkeith02,levequefort07}.

The overall fluorescence enhancement observed in  ASNOM  will be the result of a combination of these effects; experimentally, both fluorescence intensity enhancement\cite{frey04,protasenko04,gerton04,kramer02} and quenching\cite{frey04,yang00} have been reported. In addition to ASNOM, plasmon-induced fluorescence modifications are currently being investigated for a wide variety of nanostructured systems, such as metal-island films,\cite{zhang07} nanoparticle arrays,\cite{gerber07,schatz03,nelayah07} and individual gold nanospheres.\cite{novotny06,kuhn06} The strong dependence of these plasmonic effects on fluorophore-metal separation has been used to increase both lateral and longitudinal resolution in confocal microscopy, with important consequences for biological imaging.\cite{alschinger03}

Raman spectroscopy provides an enticing alternative to fluorescence, as it has many potential advantages and it is able to provide
high-resolution information with chemical specificity. As such, the technique is finding increasing application in biological
systems, as it allows rapid diagnosis using spectral deconvolution techniques such as principal component analysis.\cite{hanlon00} However, the
very weak intrinsic cross-section of the Raman scattering process (some 14 orders of magnitude less than fluorescence cross-sections) makes it
impractical for many biological applications that require low laser powers and short integration times. This Raman signal can be enhanced by
many orders of magnitude via surface enhanced Raman scattering (SERS) from metal nanoparticles.\cite{nie97} Devices that
utilize this effect can act as ultra-sensitive biosensors, with diverse applications in all areas of pathogen detection.\cite{zhang05}

Here, we report the results of investigations into the optical properties of fluorophores in the vicinity of noble metal nanotips. ASNOM imaging of  individual clusters of quantum dots indicates a strong distance-dependent enhancement of the quantum dot fluorescence due to competing radiative and non-radiative decay channels; this enhancement results in a simultaneous increase in optical resolution to approximately 60 nm.  These results are consistent with Finite Difference Time Domain (FDTD) simulations, which show a large enhancement of the electric field at the apex of a metal tip.
Periodic arrays of silver nanotips have been studied using fluorescence lifetime imaging microscopy (FLIM); we have observed localized enhancements in the emission intensity, and additional fluorescence lifetime components from nearby fluorophores.  By correlating the initial emission intensity and lifetime resulting from these modified decay channels, we attribute these enhancements to a greater photon recycling rate due to coupling with surface plasmons.   High resolution spatial maps of Raman scattering show an enhancement in signal over the nanotips of more than 5 orders of magnitude.

\section{Fluorescence and Resolution Enhancement in ASNOM}

\subsection{Field Enhancement}
\label{sec:sim}

\begin{figure*}[]
\begin{center}
\includegraphics[width=16cm]{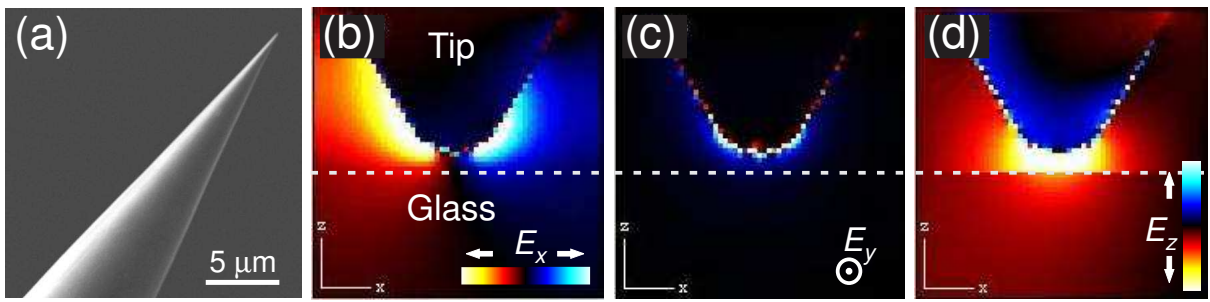}
\end{center}
\caption{ (a) SEM image of an etched Au tip, indicating an apex of $\sim$20 nm. (b-d) Respectively, x, y, and z components of the electric field vector calculated at the peak amplitude of the plasmon resonance. The blue (red) scale indicates the
field pointing along the positive (negative) direction of the corresponding axis, and the brightness indicates the relative magnitude. The computational problem space is 120 nm in each dimension.}
\label{tipsim}
\end{figure*}

Simulations of the interaction between a metal tip and a light pulse have been made using the Finite Difference Time Domain (FDTD) method.\cite{richards03} The geometry considered for the tip–surface system was based on Scanning Electron Microscope (SEM)(FEI Quanta FEG) images of etched gold tips, as shown in Fig.\ \ref{tipsim}(a): the tip was modeled as a metal cone with an end radius of 20 nm,  4 nm above a glass substrate. This system was subjected to a Gaussian derivative light pulse, polarized parallel to the x-z plane of incidence. The time-dependent electric and magnetic fields were then calculated using discrete Maxwell equations. Full details are given elsewhere.\cite{festy04}

The incident pulse excites a localized surface plasmon resonance in the tip: the magnitude of the x, y, and z components of the electric field around the tip are shown in Figs. \ref{tipsim}(b-d), respectively, at the peak amplitude of the plasmon resonance. The blue scale corresponds to the field pointing along the positive axis direction, and the red scale to
the field pointing along the negative direction. Of particular interest is the large enhancement in electric field along the z-axis between the tip apex and substrate, as shown in Fig.
\ref{tipsim}(d). These strong electric fields can lead to a greatly increased excitation and emission rate for fluorophores in the vicinity of
the tip,\cite{huang06} as well as enhancements in the Raman signal of more than seven orders of magnitude.\cite{festy04} The experimental
verification of these effects is discussed below.

\subsection{Experimental Details}

The ASNOM system is based on an inverted confocal optical microscope in which incident laser light (0.5 $\mu$W 488 nm or 532 nm) is focused to a
diffraction-limited spot on the sample surface using a high numerical aperture (NA) lens (Nikon, 100$\times$ oil, 1.45 NA). This high NA lens leads to
some total internal reflection at the glass-air interface, producing a strong evanescent field component at the sample surface. The same lens is
used to collect the fluorescence, which passes through a long-pass laser-rejection filter and a 650 nm bandpass filter, and the signal is
detected with a photomultiplier tube. Tips were prepared by electrochemically etching a 0.1 mm diameter gold wire (Aldrich) in a 1:1 mixture of fuming HCl (Fluka) and purified water. A typical etched tip used for ASNOM imaging is shown in Fig.\ \ref{tipsim}(a). The tip was positioned in the laser focus a few nanometers
above the sample using a shear-force feedback mechanism, and fluorescence images were acquired by scanning a sample through the laser spot. More
details are given elsewhere.\cite{huang05}

\subsection{Resolution Enhancement}

\begin{figure*}[]
\begin{center}
\includegraphics[width=13cm]{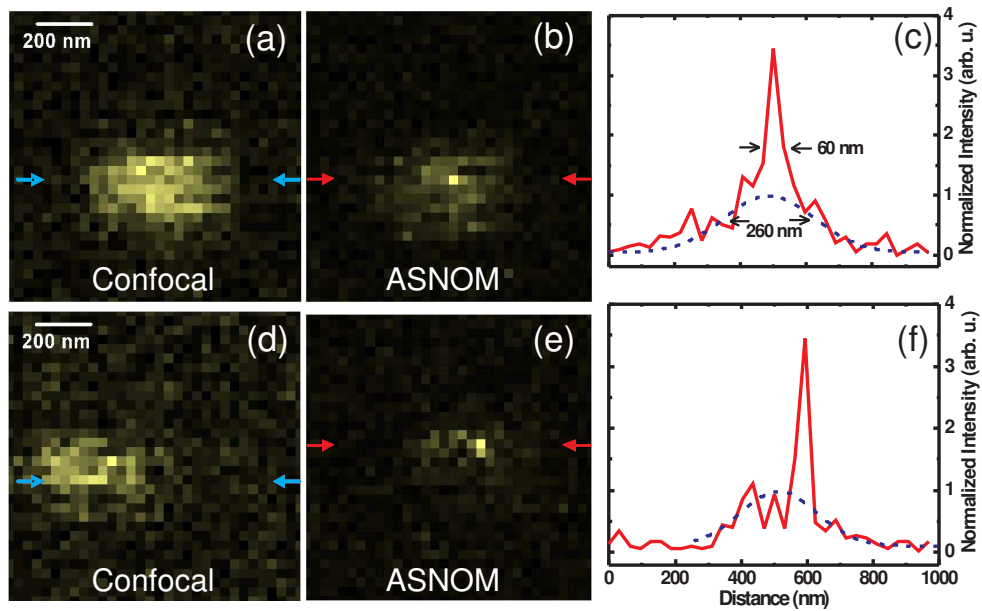}
\end{center}
\caption{ $1\times1$ $\mu$m scans of two different QD clusters, with and without a tip present: (a), (d) Confocal fluorescence images
with tip retracted. (b), (e) Corresponding ASNOM fluorescence images with a tip in feedback over the dot. In each image the intensity scale is
normalized to the maximum signal. (c), (f) Fluorescence line profiles of the confocal image (dashed line, Gaussian fit) and ASNOM image (solid
line) along the lines indicated by arrows in the corresponding images; lines have been normalized to the maximum confocal intensity. The line
profile of the confocal image in (f) is re-centered 250 nm with the ASNOM fluorescence line profile.} \label{NF1}
\end{figure*}

To study the resolution and fluorescence enhancement effects in ASNOM, small clusters of quantum dots (QDs) were investigated: a dilute solution
of CdSe/ZnS QDs (Invitrogen QDot 655) was spin coated on a clean glass coverslip in a $\sim$10 nm layer of Poly(methyl methacrylate) (PMMA)(Sigma Aldrich). This polymer layer acts a matrix to prevent dots attaching to the tip during scanning, and also reduces the effects of fluorescence blinking.

Figure \ref{NF1} shows  $1\times1$ $\mu$m  fluorescence images of two different QD clusters with and without a tip present. Figures \ref{NF1}(a)
and (d) are confocal scans taken without a tip: Gaussian fits to intensity profiles are shown as dashed lines in Figs. \ref{NF1}(c) and (d),
which indicate a resolution of about 260 nm consistent with the size of the diffraction-limited laser spot. The random fluctuations in the
signal intensity in the confocal images are due to the fluorescence intermittency or 'blinking' of the dots.\cite{nirmal96} A gold tip was then
brought within a few nanometers of the sample surface, and the same $1\times1$ $\mu$m area around the QD cluster was scanned again. The resulting
intensity maps are shown in Figs. \ref{NF1}(b) and (c), corresponding to the confocal scans in (a) and (d), respectively. In each image the
intensity scale is normalized to the maximum signal. The QD cluster has been re-centered by 250 nm between the scans shown in Figs. \ref{NF1}(d)
and (e). Line profiles through the ASNOM scans, indicated by arrows, are shown in Figs. \ref{NF1}(c) and (f); the lines have been normalized to
the maximum intensity observed in the corresponding confocal scans.

The intensity profiles in Figs. \ref{NF1}(c) and (f) show two dramatic effects: For both clusters there is an improvement in the spatial
resolution of the fluorescence image when the gold tip is in close proximity to the sample surface. The peak width in the ASNOM images is
$\sim$60 nm, a four-fold reduction over that of the diffraction limited signals observed in the confocal images. Furthermore, there is an
approximate four-fold increase in the maximum signal detected. Closer inspection of the ASNOM images in  Figs. \ref{NF1}(b) and (e), and their
respective intensity profiles, indicates that the strong sharp peak sits on a weak background signal with the same intensity and size as the
diffraction limited scans in \ref{NF1}(a) and (b). This is consistent with strong enhancement of the QD fluorescence only occurring when the QD
cluster is directly beneath the tip apex where the electric field is maximum, as shown in Fig.\ \ref{tipsim}(d).

There are two main causes of fluorescence enhancement from molecules close to a metal tip: The large increase in the local field intensity around the tip, discussed in Sec.\ \ref{sec:sim}, results in an increase in a molecule's excitation rate. The excited-state molecular dipole can also couple with surface plasmon electrons in the metal creating an additional radiative decay channel. These effects are investigated in more detail for arrays of nanotips, as discussed in Sec.\ \ref{Sec:arrays} below.

\subsection{Distance-dependent Fluorescence Enhancement}

\begin{figure*}[]
\begin{center}
\includegraphics[width=14cm]{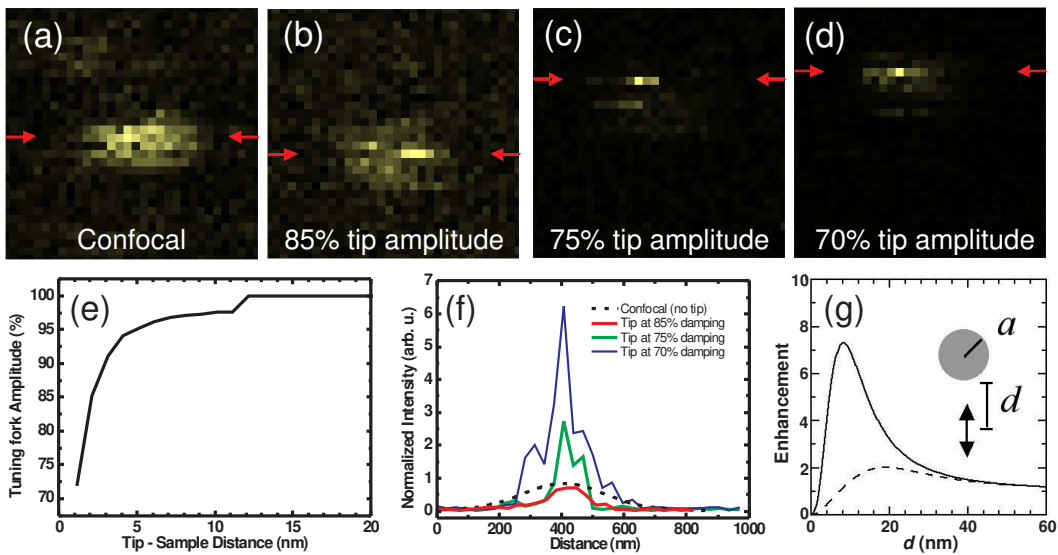}
\end{center}
\caption{ $1\times1\mu$m scans of a small QD cluster, with decreasing tip-sample separation: (a) Confocal fluorescence image with the tip
retracted 3 $\mu$m away from the sample surface. (b - d) ASNOM fluorescence image of the same cluster with the tip at $85\%$, $75\%$ \& $70\%$
amplitude damping, respectively. (e) Typical approach curve, showing the tuning fork amplitude as a function of tip-sample distance. (f)
Intensity profiles along a line indicated by arrows for each scan, normalized to the maximum confocal intensity (dashed line).  (g) Fluorescence
enhancement calculated for a dipole a distance \emph{d} from a gold sphere of radius $a=20$ nm. The solid line is for a quantum efficiency $Q =
0.1$ and the dashed line is for $Q = 1$.} \label{NFSeq}
\end{figure*}

To investigate the tip-enhanced fluorescence effects in more detail, the measurements above were repeated with the tip at different heights
above a small cluster of QDs. Figure \ref{NFSeq}(a) is a confocal scan of the dots, taken with the tip retracted far from the sample. The tip
was then gradually approached towards the sample surface, by monitoring the damping of the tuning fork used for feedback control. Subsequent
scans at $85\%$, $75\%$ and $70\%$ amplitude are shown in Figs.\ \ref{NFSeq}(b-d), respectively. Figure \ref{NFSeq}(e)
shows how the relative tuning fork oscillation amplitude varies as a function of tip-sample separation, for a tip approaching a
surface.\cite{huang07}

Intensity profiles for each ASNOM scan are shown in Fig.\ \ref{NFSeq}(f), along lines indicated by arrows in each scan. The intensity of each
line profile has been normalized to that of the confocal scan with tip retracted, shown by a dashed line in (f). The first scan
taken with the tip present at $85\%$ amplitude shows no increase in fluorescence intensity from the QDs, relative to the confocal scan; however,
there is an increase in resolution and the confocal 'background' is much lower than in Fig.\ \ref{NFSeq}(a). This is consistent with an overall
reduction in fluorescence intensity from the QDs due to bleaching.

Despite this bleaching, a further reduction in tip-sample distance produces a significant increase in QD fluorescence signal, with a
corresponding increase in resolution, as shown in Figs.\ \ref{NFSeq}(c) and (d) and the line profiles in (f). The small vertical shift between
the scans at $85\%$ and $75\%$ is due to sample drift. A six-fold increase in the peak signal is measured at $70\%$ damping, relative to the
original confocal image; the enhancement factor is even greater if the effects of bleaching are considered. Figure \ref{NFSeq}(e) shows that a
reduction in the tuning fork amplitude from $85\%$ to $70\%$ typically corresponds to a decrease in tip-sample distance of only a few
nanometers.

These results are consistent with calculations of the distance-dependent fluorescence enhancement near to a gold sphere.\cite{huang06} These
calculations are summarized in Fig.\ \ref{NFSeq}(g), which shows the enhancement in fluorescence from a dipole a distance \emph{d} from a gold
sphere of radius $a=20$ nm. The solid line is for a quantum efficiency $Q = 0.1$ and the dashed line is for $Q = 1$. It should be noted that the
QDs used here have a polymer / biomolecule shell, which means that for the six-fold fluorescence enhancement observed at $70\%$ amplitude, the
actual separation between the tip and QD core is still $\sim$10 nm. This is in excellent agreement with Fig.\ \ref{NFSeq}(g), which also
predicts a large reduction in enhancement factor for an increase in tip-sample separation of only a few nanometers.

\section{Nanotip Arrays}
\label{Sec:arrays}

In the previous section we presented theoretical and experimental results highlighting the modifications to the optical properties of fluorophores in close proximity to a sharp metal tip. These effects have also been investigated for  arrays of self-assembled triangular nanotips, which exhibit many similar properties to etched metal tips, under optical excitation.

\subsection{Experimental Details}

\begin{figure}[tb]
\begin{center}
\includegraphics[width=8cm]{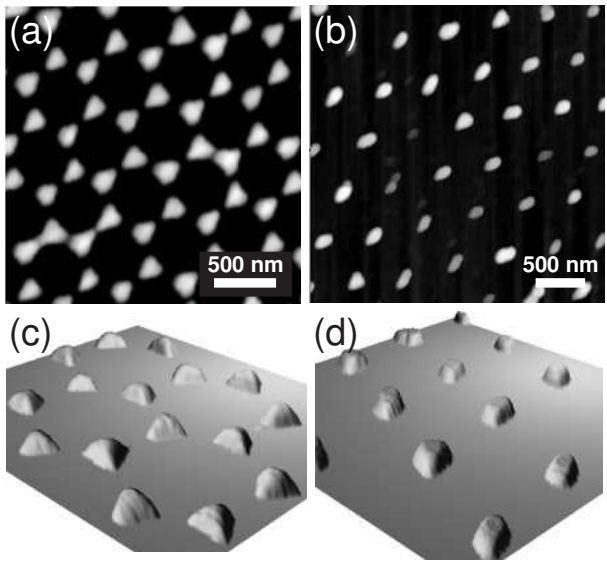}
\end{center}
\caption{ AFM images of nanotip arrays: (a) Triangular nanotips formed by Ag deposition on a single layer of latex spheres.  (b) Nanotips formed using a double layer of spheres. (c) and (d) 3-dimensional representations of (a) and (b), respectively. } \label{afm}
\end{figure}

Glass coverslips were cleaned in piranha solution for an hour to remove organic matter and hydroxylate the surface. 500 nm diameter latex spheres (Invitrogen) were diluted in deionized water and drop-cast onto a coverslip, and left to dry; as the water evaporates the spheres assemble into a monolayer / bilayer close-packed lattice.\cite{vanduyne01} A 0.5 nm layer of chromium was then deposited onto the slide by thermal
evaporation in a vacuum chamber; this increases the adhesion of silver which was subsequently deposited to a thickness of 25 nm. The latex
spheres were removed by sonication in chloroform, leaving a continuous periodic array of Ag nanotips over a typical area of 20 mm$^2$.
Figures \ref{afm}(a) and (c) show atomic force microscopy (AFM) images of nanotips formed from a single layer of latex spheres. Figure \ref{afm}(c) is a 3 dimensional representation showing individual particles within the array, which have a base length of $\sim$120 nm and height of 25 nm. In some regions of the sample, different structures form due to Ag deposition on double layers of spheres,\cite{vanduyne01} as shown in Figs.\ \ref{afm}(b) and (d).

Rhodamine 6G (R6G) dye (Sigma Aldrich) was deposited onto the nanotips by vacuum sublimation to produce a uniform coverage of sub-monolayer thickness. The
sample was prepared for high resolution optical measurements by applying a thin layer of index-matching polymer solution (Mowiol 4-88, Sigma Aldrich) to a clean
coverslip and affixing this on top of the nanotip array. During this procedure the R6G was incorporated into the polymer solution, creating
a homogeneous distribution of fluorophores across the sample with a thickness of several microns. This was verified by spectroscopic analysis of
excess polymer from the edge of the slide.

Fluorescence lifetime and intensity images were obtained using a scanning confocal epifluorescence microscope  (Leica TCSP2, 100x oil objective,
1.4 NA) with a synchronous time correlated single photon counting module (Becker and Hickl SPC-830). Excitation was with a 488 nm
continuous-wave (CW) Ar$^{+}$ laser (intensity) or 467 nm 20 MHz pulsed diode laser (lifetime), at a power far below fluorescence saturation in
both cases.

Raman spectra and images were obtained using a Renishaw spectrometer with a high precision scanning stage (Prior H101). To reduce
fluorescence background, excitation was at 752 nm using a home-made tunable CW Ti:Sa laser with the 100x oil objective.

\subsection{Fluorescence Enhancement}
\label{sec:fluo}

\begin{figure}[tb]
\begin{center}
\includegraphics[width=8cm]{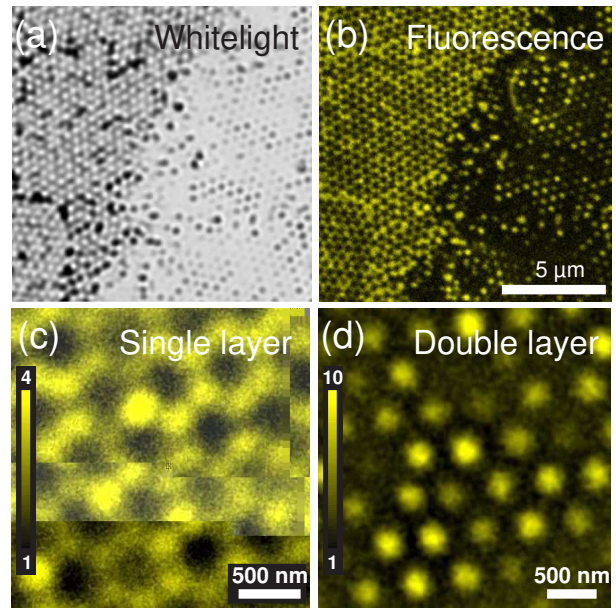}
\end{center}
\caption{ (a) Transmitted white light image of a boundary region between single layer (left) and double layer (right) nanotips. (b)
Corresponding confocal  fluorescence intensity map of R6G, for the same region. (c) and (d) High resolution fluorescence maps from single and
double layer regions similar to Fig.\ \ref{afm}(a) and (b), respectively.} \label{fluorescence}
\end{figure}

Figure \ref{fluorescence}(a) is the transmitted white light image of a boundary region between  nanoparticles formed by a single layer of latex
spheres (left) and a double layer (right). The corresponding confocal R6G fluorescence signal detected from this region is shown in Fig.\
\ref{fluorescence}(b). High resolution scans from this area are shown in Figs.\ \ref{fluorescence}(c) and (d) for the single and double layer
nanotips, respectively. These maps show highly localized fluorescence enhancement from molecules close to nanotips. Spectral analysis
has verified that the emission originates from R6G and is not caused by photoluminescence from the silver nanotips or scattered laser
light. The R6G fluorescence enhancement measured from the nanotips, relative to that from the glass, is approximately four-fold and ten-fold
for (c) and (d), respectively. The actual enhancement in the vicinity of the nanotips will be much larger than that measured: Fig.\ \ref{NFSeq}(g) shows that significant enhancement only occurs over a very small range of metal-fluorophore distances ($\sim$20 nm),\cite{novotny06} hence there is a large unmodified fluorescence background from the other fluorophores in the excited confocal volume. This is discussed further below.

\subsection{Lifetime Modification}
\label{sec:lifetime}

For CW measurements it is not possible to separate the contributions to the fluorescence enhancement arising from the modified excitation and
decay channels; however, this is not the case with time-resolved measurements. The quantum efficiency $Q$ gives the probability of an excited
state molecule decaying to a lower state by photon emission. In free-space $Q_{0}=\Gamma \tau_{0}$, where $\Gamma$ is the radiative emission
rate, $\tau_{0} = (\Gamma + k_{\mathrm{nr}})^{-1}$ is the total lifetime of the molecule, and $k_{\mathrm{nr}}$  is the sum of the non-radiative
decay rates. The presence of a metal creates an additional decay channel $\Gamma_{\mathrm{m}}$ for an excited molecule, so the total radiative
decay rate becomes $\Gamma^{\prime}=\Gamma+\Gamma_{\mathrm{m}}$. The modified fluorescence quantum yield $Q_{\mathrm{m}}$ and lifetime
$\tau_{\mathrm{m}}$ are then related by $Q_{\mathrm{m}}=\Gamma^{\prime}\tau_{\mathrm{m}}$, and $\tau_{\mathrm{m}} = (\Gamma^{\prime}+
k_{\mathrm{nr}}^{\prime})^{-1}$, where $k_{\mathrm{nr}}^{\prime}$ is the modified total non-radiative decay rate.

\begin{figure*}[tb]
\begin{center}
\includegraphics[width=14cm]{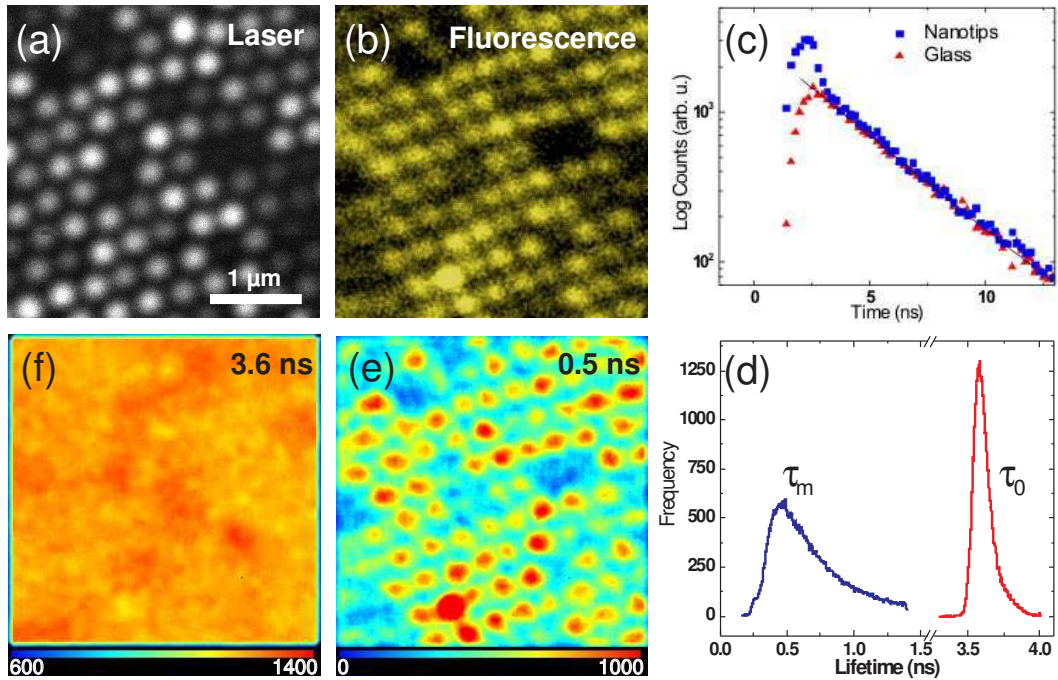}
\end{center}
\caption{Fluorescence lifetime analysis for a region of double-layer nanotips:  (a) Reflected laser intensity and (b) integrated
fluorescence, from a small region similar to Fig.\ 2(d). (c) Raw decay transients from R6G on glass and nanotips; the straight line is a
linear fit to the glass data. (d) Histogram of lifetime components ($\tau_{\mathrm{m}}$, $\tau_{0}$) obtained from biexponential pixel fits of
the scan area. (e) Spatial map of the preexponential intensity $\alpha_{\mathrm{m}}$ for the 0.5 ns $\tau_{\mathrm{m}}$ component. (f) as (e) mapping $\alpha_{0}$ for the 3.6 ns $\tau_{0}$ component, showing homogeneity across the region. Note the difference in the intensity scales.} \label{lifetime}
\end{figure*}

The effects of Ag nanotips on the R6G fluorescence lifetime were investigated for a region of double-layer structures: Figures \ref{lifetime}(a) and (b) show the reflected laser and fluorescence intensity maps, respectively, for a scan area similar to that of Fig.\ \ref{fluorescence}(d). After excitation by a short laser pulse, the time
evolution of the fluorescence intensity of a free-space single molecule is $I(t)= \alpha e^{-t /\tau_{0}}$. The
spatially integrated decay transients from large areas of glass and nanotips are shown in Fig.\ \ref{lifetime}(c). In both cases there is
an identical long-lived component originating from unmodified fluorophores in the polymer layer; this has a monoexponential decay $>$3 ns and
implies that there are no significant concentration-induced non-radiative channels.\cite{leitner85} A FLIM map of this region was acquired for a
30 min integration period. Biexponential fits were applied to small bins of pixels to give spatially resolved maps of constituent lifetimes $\tau_{i}$ and their corresponding intensities $\alpha_{i}$. A histogram of the extracted lifetime components is shown in Fig.\ \ref{lifetime}(d): this comprises a sharply peaked
component $\tau_{0}$ from the unperturbed fluorophores, and a broader fast component $\tau_{\mathrm{m}}$ arising from fluorophores with modified
decay rates. Figures \ref{lifetime}(e) and (f) show maps of the preexponential intensities $\alpha_{\mathrm{m}}$ and $\alpha_{0}$ for the two modal lifetime components; the 0.5 ns
component is localized (resolution limited) around the nanotips, whereas the 3.6 ns component has a uniform intensity (standard deviation
$< 2 \%$) consistent with the homogeneous coverage of fluorophores.\cite{rm07}

For this sample, the R6G (in polymer) layer thickness is comparable to the axial confocal excitation depth; hence, fluorophores at different
distances from the nanotips will contribute differently to the total signal measured. Molecules in very close proximity to a metal show
strong fluorescence quenching due to dominant non-radiative energy transfer.\cite{dulkeith02} We have taken measurements on samples without the
additional polymer layer, where the R6G molecules form a monolayer; these measurements show almost complete quenching of the R6G fluorescence over the Ag nanotips, in agreement with the calculations shown in Fig.\ \ref{NFSeq}(g). This strong distance-dependent weighting means that, for the current sample, the \emph{enhanced} fluorescence with a lifetime $\tau_{\mathrm{m}}$ is from a thin shell of molecules $\sim$10 nm away from the nanotips: at this distance, the competing quenching and enhancement mechanisms result in a maximum net emission intensity.

An aqueous R6G molecule has $Q=0.95$; hence, we assume that fluorophores showing a maximally enhanced emission intensity have $Q_{\mathrm{m}}
\approx 1$, so $\Gamma^{\prime} \approx \tau_{\mathrm{m}}^{-1}$.  In CW measurements, the nanotip-induced enhancement in integrated intensity cannot be accounted for by an increase in $Q$. Instead it originates primarily from a greatly increased photon recycling rate for fluorophores with a strong radiative coupling to surface plasmons.

\subsection{Surfaced Enhanced Raman Scattering}
\label{sec:sers}

\begin{figure}[]
\begin{center}
\includegraphics[width=8cm]{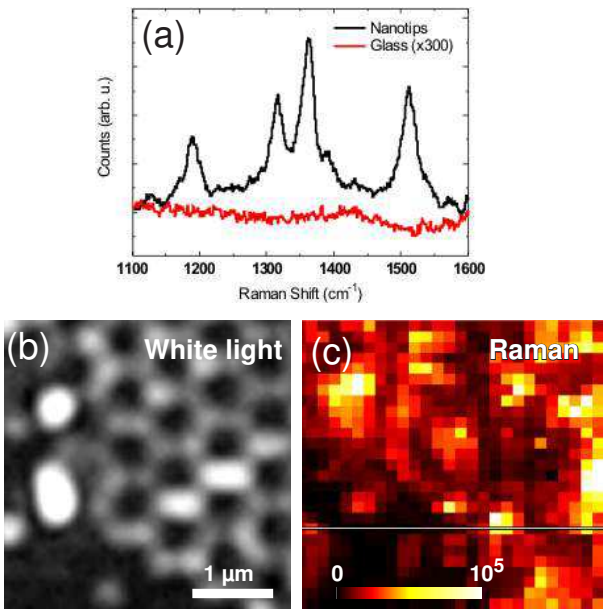}
\end{center}
\caption{ (a) Raman spectra from R6G on nanotips and glass. (b) Reflected white light image of nanotip array. (c) Intensity map of the
1500 cm$^{-1}$ Raman line, from a region similar to (b).} \label{raman}
\end{figure}

Figure \ref{raman}(a) shows the Raman spectra from R6G on a region of single-layer nanotips (2 s acquisition time) and from R6G on glass (10 min
acquisition time). As no signal is obtained from the glass region, it is not possible to calculate an absolute value for the enhancement factor;
however, a lower limit of $10^5$ can be estimated using the noise level in the glass spectrum. This has been verified by obtaining high
resolution spatial maps of the Raman signal over the nanotips. Figure \ref{raman}(c) shows the integrated intensity under the 1500
cm$^{-1}$ Raman line, for a region of nanotips similar to \ref{raman}(b). The huge localized enhancement in Raman signal is due to the
increase in the local electric field intensity around an individual nanotip: the total SERS enhancement
has an $E^4$ dependence on electric field due to enhancement of the incident laser field and enhancement of the emission field at the Raman
frequency.\cite{kneipp02} These results are consistent with simulations of the electric field around a single sharp metal tip, presented in Fig.\ \ref{tipsim}; this is reported in detail elsewhere,\cite{festy04,demming05} where theoretical enhancements of up to nine orders of magnitude are obtained.

The surface plasmon resonance of an individual nanotip is highly sensitive to small changes in its size, shape, and local environment, as well as electromagnetic coupling between nearby tips.\cite{klar98,jin01,mock02} These small variations lead to localized differences in the field enhancement, resulting in the creation of optical ``hot spots", as seen in Fig.\ \ref{raman}(c).\cite{gresillon99}

\section{Conclusion}

Here, we have presented an investigation into modifications to the optical properties of fluorophores, in the presence of metal nanotips.
Simulations of the electric field around an irradiated tip show a large highly localized increase in field between the tip and sample. Experimentally, we observe an order of magnitude enhancement in fluorescence from quantum dots, resulting in a simultaneous increase in resolution many times greater than the far-field diffraction limit. These effects have been further investigated for ordered arrays of Ag nanotips. FLIM maps of R6G fluorescence show an additional plasmon-induced radiative decay channel for a thin shell of molecules around individual nanotips. This
leads to an order of magnitude increase in ensemble photon recycling rate with a proportionate enhancement in fluorescence intensity. The strong
distance dependence of these plasmonic effects suggests that tip enhanced fluorescence may offer a means of selectively mapping the location of specific target molecules, such as fluorophore-tagged proteins in cell membranes. Furthermore, spatially resolved maps
of Raman scattering from R6G show localized enhancement from individual nanotips of more than five orders of magnitude. This has important
implications for the development of ultra-sensitive biosensors that require a high degree of chemical specificity.

\section{Acknowledgements}

The authors would like to thank K. Suhling for helpful discussions. This work was supported by the EPSRC (UK).

\end{document}